\documentclass{article}
\usepackage[a4paper, total={6in, 8in}]{geometry}
\usepackage[utf8]{inputenc}
\usepackage{graphicx}
\usepackage{dirtytalk}
\usepackage{amssymb}
\usepackage{amsfonts}
\usepackage{amsmath}
\usepackage[T1]{fontenc}

\begin{document}
\vspace*{0.35in}

% title goes here:
\begin{flushleft}
{\Large
\textbf\newline{Generation of 0.7 mJ multicycle 15 THz radiation by phase-matched optical rectification in lithium niobate}
}
\newline
% authors go here:
\\
Dogeun Jang\textsuperscript{1,4},
Jae Hee Sung\textsuperscript{2,3,4},
Seong Ku Lee\textsuperscript{2,3},
Chul Kang\textsuperscript{2,5},
Ki-Yong Kim\textsuperscript{1,6}
\\
\bigskip
\bf{1} Institute for Research in Electronics and Applied Physics, University of Maryland, College Park, MD 20742
\\
\bf{2} Advanced Photonics Research Institute, Gwangju Institute of Science and Technology, 123 Cheomdangwagi-ro, Oryong-dong, Buk-gu, Gwangju, 61005, South Korea
\\
\bf{3} Center for Relativistic Laser Science, Institute for Basic Science, 123 Cheomdangwagi-ro, Buk-gu, Gwangju, 61005, South Korea
\\
\bf{4} These authors contributed equally to this work
\\
\bf{5} e-mail: iron74@gist.ac.kr
\\
\bf{6} e-mail: kykim@umd.edu
\\
\bigskip
%* kykim@umd.edu, iron74@gist.ac.kr

\end{flushleft}

%\maketitle

%\title{\textbf{Generation of 0.7 mJ multicycle 15 THz radiation by phase-matched optical rectification in lithium niobate} \\ \textit{\normalsize Institute for Research in Electronics and Applied Physics, University of Maryland, MD 20742, USA}}

%\author{\normalsize Dogeun Jang, Jae Hee Sung, Seong Ku Lee, Chul Kang, and K. Y. Kim}

%\date{}

\begin{abstract}
We demonstrate efficient multicycle terahertz pulse generation at 14.6 THz from large-area lithium niobate crystals by using high-energy (up to 2 J) femtosecond Ti:sapphire laser pulses. Such terahertz radiation is produced by phase-matched optical rectification in lithium niobate. Experimentally, we achieve maximal terahertz energy of 0.71 mJ with conversion efficiency of $\sim$0.04\%. We also find that the effective interaction length for optimal terahertz conversion is fundamentally limited by terahertz absorption, laser pulse stretching by material dispersion, and terahertz-induced nonlinear cascading effects on the driving laser pulse.  
\end{abstract}

\section*{Introduction}
Strong terahertz radiation is essential for studies of terahertz-driven linear and nonlinear phenomena in various novel materials \cite{ferguson2002materials, schmuttenmaer2004exploring}, molecular alignment \cite{fleischer2011molecular, kampfrath2013resonant, egodapitiya2014terahertz}, high harmonic generation \cite{balogh2011single, schubert2014sub, kovacs2012quasi}, to name a few. Recently, various terahertz generation schemes have been proposed and developed. Among those, optical rectification (OR) in the $\chi^{(2)}$ nonlinear crystal is considered one of the promising methods to generate high-power terahertz radiation \cite{fulop2014efficient}. For instance, the lithium niobate (LN) crystal has emerged as a strong terahertz emitter due to its high nonlinearity \cite{hebling2004tunable} and high damage threshold \cite{nakamura2002optical}. Also its large bandgap ($\sim$4 eV) allows convenient optical pumping at 800 nm without a concern of two photon absorption.

In particular, prism-shaped LN crystals are widely used  for efficient terahertz generation with a tilted pulse front (TPF) technique \cite{hebling2004tunable, fulop2014efficient, wu2018highly}. This allows optical-terahertz phase matching in which the group velocity of the laser pulse is matched with the phase velocity of the terahertz wave \cite{hebling2002velocity}. Recently, a LN prism as large as 68 mm $\times$ 68 mm $\times$ 64 mm was used to produce single-cycle terahertz pulses with energy up to 0.2 mJ per pulse  \cite{wu2018highly}. The prism-based geometry, however, gives rise to nonuniform pump interaction inside the crystal due to its large pump tilt angle ($63^{\circ}$ for excitation at 800 nm). This not only results in poor terahertz beam quality but also makes it impractical to upscale the excitation setup for higher terahertz output energy. 

Lately, a planar LN wafer with large-aperture (75-mm diameter) has been used to generate high-energy (0.19 mJ) terahertz pulses at <5 THz \cite{jang2019scalable}. This planar geometry permits convenient energy-upscalable terahertz generation. In this scheme, most of terahertz energy is produced from the front and rear surfaces of LN within one coherence length (50 $\mu$m at 1 THz) \cite{jang2019scalable}. However, no phase matching is generally satisfied at low terahertz frequencies (<5 THz). Surprisingly, planar LN turns out to allow phase-matched terahertz generation at high frequencies, for example, $\sim$15 THz when pumped at 800 nm \cite{jang2019hidden}. The radiation is also characterized with multicycle narrowband emission \cite{jang2019hidden} unlike prism-based LN schemes.

In  this  paper,  we  present  an experimental demonstration of high-energy multicycle terahertz radiation at 14.6 THz by using large-area planar LN crystals. Here we examine the effects of incident laser energy/chirp and LN properties (thickness and MgO doping) on the phase-matched terahertz generation to determine the optimal conditions. We also analyze the generation process via numerical simulations.

\section*{Experimental Setup}
\begin{figure*}[b!]
\centering
\includegraphics[width=14cm]{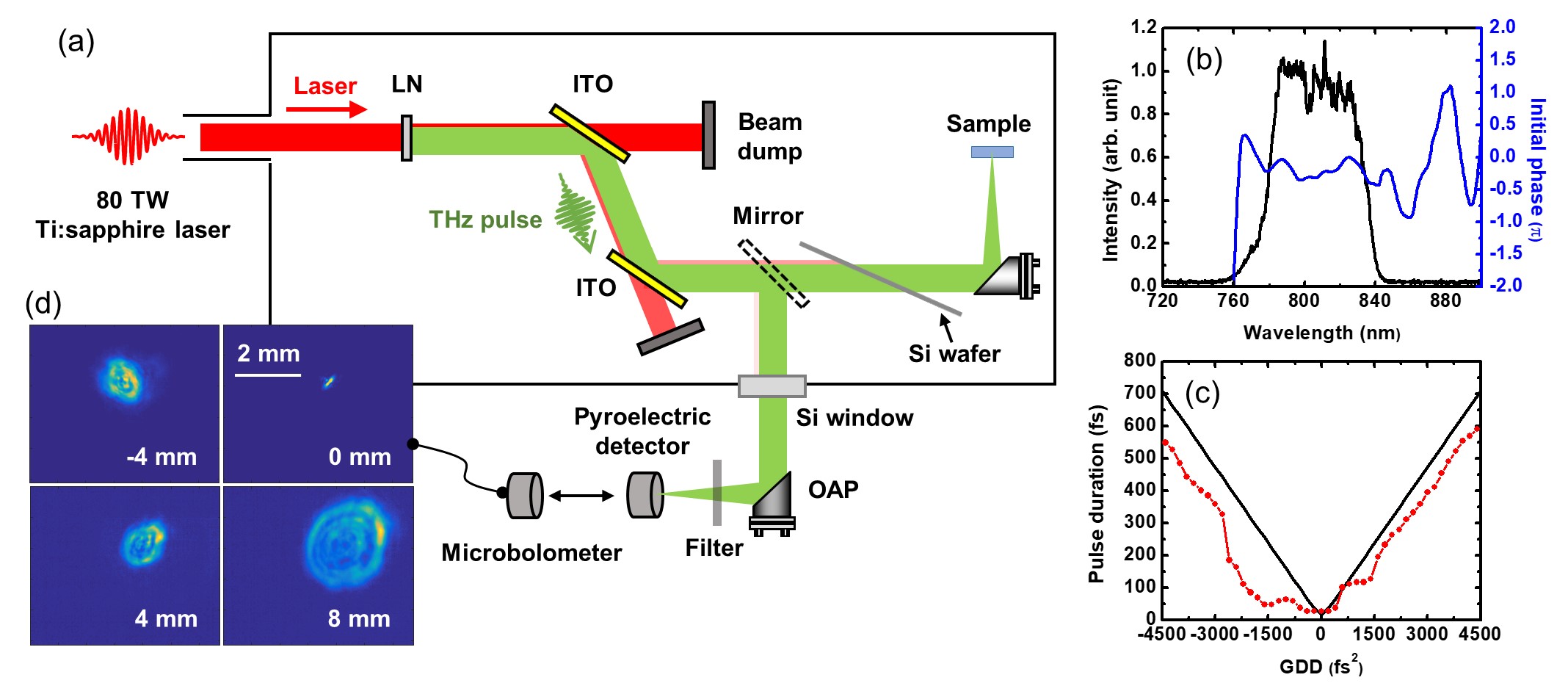}
\caption{(a) Experimental setup for multicycle terahertz pulse generation from a bulk cLN. (b) Input laser spectral intensity (black line) and phase (blue line) with zero GDD applied. (c) Input laser pulse duration in FWHM estimated from a Gaussian assumption (black line) or  Fourier transform (red line with circles). (d) Terahertz beam profiles at -4, 0, 4, and 8 mm from the focal point of the OAP mirror.}
\label{fig1}
\end{figure*}

Our experiment was performed with a Ti:Sapphire laser system capable of delivering 25-fs, 2-J pulses in single-shot mode \cite{jang2019scalable}. A schematic diagram of our experimental setup is shown in Fig. \ref{fig1}(a). The incident laser pulse is p-polarized, parallel to the extraordinary axis of the LN crystal for maximal terahertz generation. In our experiments, three different congruent LN (cLN) wafers were used: x-cut, 0.5-mm-thick (5 mol\% MgO-doped and undoped) and y36-cut, 35-$\mu$m-thick undoped cLNs. The wafer diameter is 2" (MgO doped) or 3" (undoped). The generated terahertz pulse is decoupled from the co-propagating laser pulse by two optical windows coated with 250-$\mu$m-thick tin-doped indium oxide (ITO) layers. Tilted at Brewster’s angle of $56^{\circ}$, the decouplers provide overall reflection of 0.0001\% at 800 nm and 64\% at 15 THz \cite{jang2019scalable}.
The emitted terahertz radiation is transmitted through a 675-$\mu$m-thick Si wafer tilted at Brewster's angle of $66^{\circ}$ (70\% transmission at 15 THz) and then focused onto a sample for further studies. For characterization, the emitted terahertz radiation is intercepted by a movable metallic mirror and brought into air through a 10-mm-thick Si vacuum window. The radiation is then focused by a 6"-focal-length off-axis parabolic (OAP) mirror and measured by a pyroelectric detector (Gentec, THz5D-MT-BNC) along with filters including 1-mm-thick Si ($\sim$46\% transmission at 15 THz), 7.3-$\mu$m longpass ($\sim$48\%), and 15-THz bandpass ($\sim$42\%) ones. The 10-mm-thick Si vacuum window provides 14.4\% transmission at 15 THz.

Figure \ref{fig1}(b) shows the spectral phase and intensity of the input laser pulse, centered at 808 nm with a 54 nm full-width half-maximum (FWHM) bandwidth. The pulse's group delay dispersion (GDD) is controlled by an acousto-optic programmable dispersive filter (Dazzler, Fastlite), which imposes a negative or positive chirp onto the laser pulse. The GDD-dependent laser pulse duration is retrieved and plotted in Fig. \ref{fig1}(c). The red line with circles represents the pulse duration extracted from the actual spectral intensity and phase shown in Fig. \ref{fig1}(b), whereas the black solid line is obtained with a Gaussian envelope assumption with no spectral phase higher than GDD considered. Figure \ref{fig1}(d) shows terahertz beam profiles obtained with a microbolometer camera (FLIR, tau2) \cite{jang2019spectral}.

\section*{Background}
\begin{figure}[b!]
\centering
\includegraphics[width=10cm]{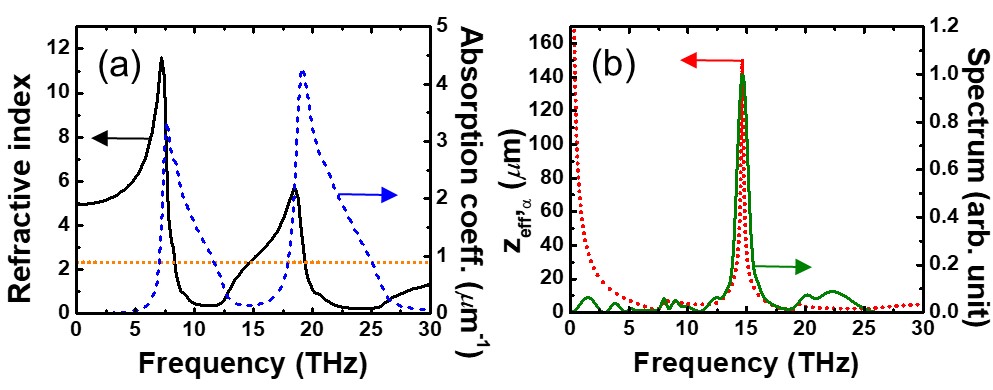}
\caption{(a) Extraordinary refractive index $n_{\text{THz}}$ (black line) and absorption coefficient $\alpha_{\text{THz}}$  (blue dotted line) of cLN, co-plotted with laser group index $n_{g} =$ 2.3 at 808 nm (orange dotted line). (b) Calculated effective interaction length $z_{\text{eff,a}}$ (red dotted line) and corresponding radiation spectrum (green solid line) expected from a 35-$\mu$m-thick cLN crystal.}
\label{fig2}
\end{figure}

In phase-matched OR, laser-to-terahertz conversion efficiency is expected to increase linearly with input laser energy. However, recent experiments and calculations have shown deviations from the linearity owing to linear and nonlinear effects in the $\chi^{(2)}$ crystal including dispersion, self-phase modulation, cascading, and stimulated Raman scattering \cite{ravi2014limitations}. These effects exhibit strong saturation or reduction behaviors in the conversion efficiency by decreasing the effective interaction length, $z_{\text{eff}}$, the distance over which the cumulative terahertz energy becomes maximal. In addition, terahertz absorption by background and free-carriers produced by multi-photon absorption in the nonlinear material is another factor that decreases $z_{\text{eff}}$ \cite{fulop2012generation}. 

The terahertz radiation we observe is characterized by phase-matched optical rectification, $n_{g} = n_{\text{THz}}$, where $n_{g}$ and $n_{\text{THz}}$ is the group and refractive index at laser and terahertz frequency, respectively. This is evident from Fig. \ref{fig2}(a) plotting the refractive index $n_{\text{THz}}$ and absorption coefficient $\alpha_{\text{THz}}$ of bulk cLN along its extraordinary axis \cite{palik1997lithium}. It shows that the condition ($n_{g} = n_{\text{THz}}$) is met at 8.3 THz, 14.6 THz, and 19.3 THz. In particular, radiation at 14.6 THz is expected to be dominant because of its relatively low absorption. From the data in Fig. \ref{fig2}(a), the effective interaction length limited by absorption, $z_{\text{eff,a}}$, is calculated via Eq. (16) in Ref. \cite{hattori2007simulation} and plotted in Fig. \ref{fig2}(b). Also plotted is the radiation spectrum expected from a 35-$\mu$m-thick cLN emitter. It is obtained from the analytical solution in Ref. \cite{hattori2007simulation} with phase-mismatching and absorption ($\alpha_{L} =$ 0.0078 $\text{cm}^{-1}$ at 808 nm \cite{schwesyg2011interaction}) effects taken into account. Both calculations predict narrow band emission at 14.6 THz.

%\begin{equation}\label{eq2}
 %\phi_s = \phi_0 + b_1(\omega-\omega_c) + b_2(\omega-\omega_c)^2 + b_3(\omega-\omega_c)^3 + ...,
%\end{equation}

\section*{Experimental Results}
\begin{figure}[b!]
\centering
\includegraphics[width=10cm]{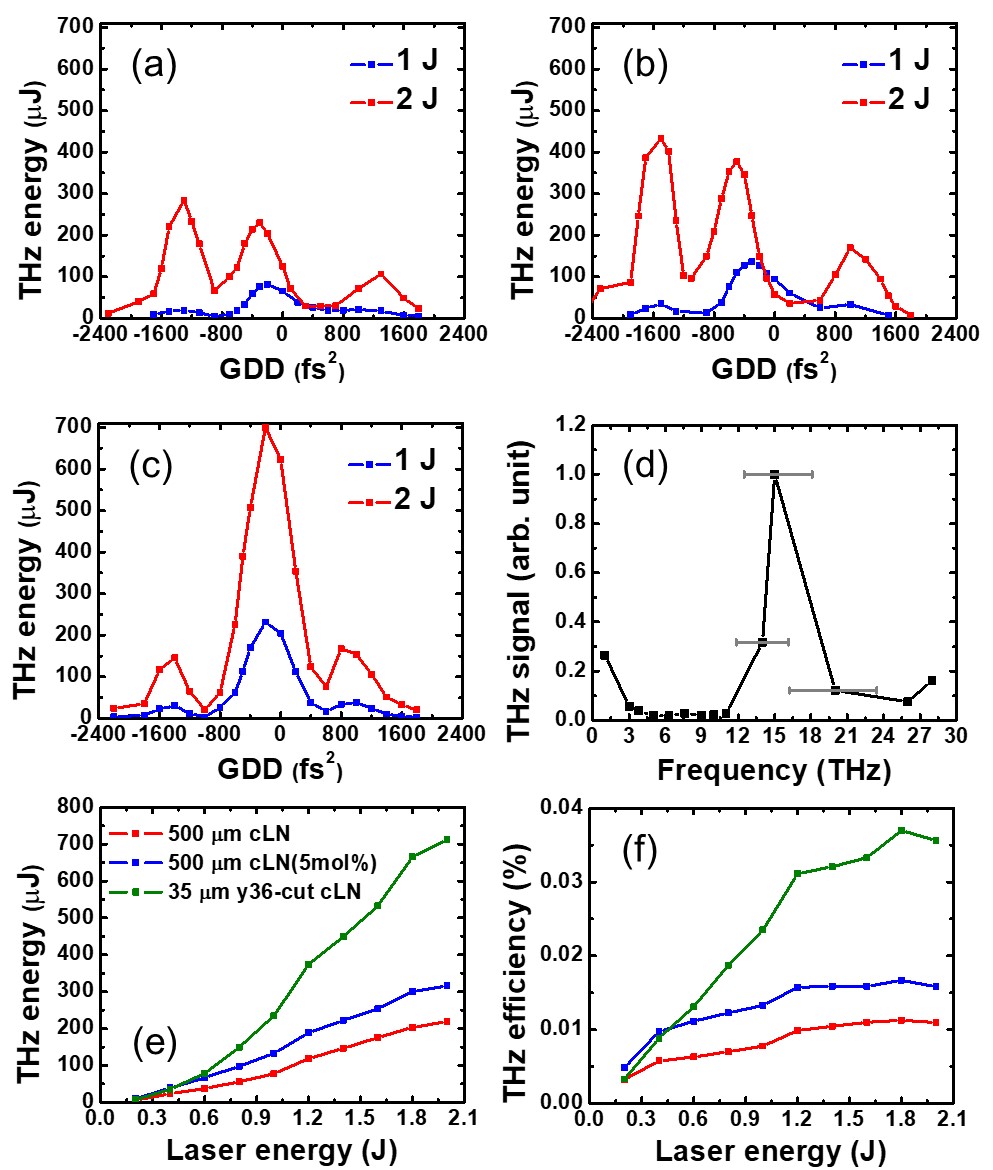}
\caption{Terahertz output energy measured as a function of laser GDD with 1 J (blue line) and 2 J (red line) energy incident on (a) x-cut, 0.5-mm-thick cLN, (b) x-cut, 0.5-mm-thick, 5 mol\% MgO-doped cLN, and (c) y36-cut, 35-$\mu$m-thick cLN. (d) Terahertz spectrum measured from the 35-$\mu$m-thick cLN. The error bar represents a half-power bandwidth at 14 THz, 15 THz, and 20 THz. (e) Terahertz output energy and (f) conversion efficiency versus input laser energy.}
\label{fig3}
\end{figure}

Our experimental results are shown in Fig. \ref{fig3}. Figures \ref{fig3}(a)–(c) plot the output terahertz energy right after the crystal for the three cLNs. In general, a slightly negative chirped laser pulse 
produces more terahertz energy compared to zero-GDD ones due to intrinsic positive material dispersion of LN. There are also additional peaks occurring at negative GDD of -1300$\sim$-1400 fs$^{2}$ and positive GDD of 800$\sim$1300 fs$^{2}$. These peaks are due to uncompensated high order dispersion (HOD) in the input laser pulse (see Fig. \ref{fig1}(c)). A proper combination of GDD and HOD can induce pulse break-ups, resulting in multiple short laser pulses and thus yielding strong terahertz outputs.

Figure \ref{fig3}(d) shows a measured terahertz spectrum. It is obtained by using 14 sets of terahertz bandpass filters in pyroelectric detection \cite{yoo2018broadband}. The spectrum is from the y36-cut cLN when excited by an unchirped laser pulse (GDD $=$ 0 fs$^{2}$) with energy of 2 J. It shows that the peak frequency is located around 15 THz, consistent with the predicted spectrum from Fig.  \ref{fig2}(b). The measured spectrum, however, appears to be not as narrow as expected. This is because the filters used at high frequencies have broadband responses and thus allow transmission at 18-25 THz as well (see Fig. \ref{fig2}(b)).

Figure \ref{fig3}(e) shows measured terahertz output energy with increasing laser energy for the three cLNs. Here the laser GDD value is fixed at -300 fs$^{2}$ for both 500-$\mu$m-thick cLNs and -200 fs$^{2}$ for the 35-$\mu$m-thick one, respectively. The corresponding laser-to-terahertz conversion efficiency is shown in Fig. \ref{fig3}(f), with a maximum value of 0.037\% achieved with the 35-$\mu$m-thick cLN. Here a 15-THz bandpass filter was used to measure terahertz energy near 15 THz only. 

Among the tested cLNs, the 35-$\mu$m one yields the highest terahertz energy (0.71 mJ). With a y36-cut, the polarization axis of the normal incident laser is $54^{\circ}$ tilted to the crystal's extraordinary axis (z-axis), which is not optimal for terahertz generation. Nonetheless, it provides the highest output energy because its thickness is the closest to the effective interaction length $z_{\text{eff}}$ among the examined cLNs. Generally, $z_{\text{eff}}$ can be enhanced by decreasing terahertz absorption with moderated MgO doping and/or cooling \cite{palfalvi2005temperature}. Because of this, the MgO-doped LN crystal yields more terahertz radiation than pure LNs as shown in Figs. \ref{fig3}(e)-(f).

\section*{Numerical Simulations}
To simulate terahertz generation from a bulk cLN, we solve 1-dimensional (1-D) coupled forward Maxwell equations (FME) for both terahertz and laser fields in the frequency domain \cite{jang2019scalable,husakou2001supercontinuum} as

\begin{equation}
\begin{split}
\frac{\partial E_{T}\left(\omega_{T}, \xi \right)}{\partial\xi}  = & -\frac{\alpha_{T}}{2}E_{T}\left(\omega_{T}, \xi \right) - jD\left(\omega_{T}\right)E_{T}\left(\omega_{T}, \xi \right) \\
& -j\frac{k_{T}d_{\text{eff}}}{n_{T}^{2}}\int\limits_{0}^{\infty} E_{L}\left(\omega_{L} + \omega_{T}, \xi \right)E_{L}^{*}\left(\omega_{L}, \xi \right) d\omega_{L}, 
\end{split}
\label{eq1}
\end{equation}
\begin{equation}
\begin{split}
\frac{\partial E_{L}\left(\omega_{L}, \xi \right)}{\partial\xi}  = & -\frac{\alpha_{L}}{2}E_{L}\left(\omega_{L}, \xi \right) - jD\left(\omega_{L}\right)E_{L}\left(\omega_{L}, \xi \right) \\
& -j\frac{k_{L}d_{\text{eff}}}{n_{L}^{2}}\int\limits_{0}^{\infty} E_{L}\left(\omega_{L} - \omega_{T}, \xi \right)E_{T}^{*}\left(\omega_{T}, \xi \right) d\omega_{T} \\
& -j\frac{k_{L}c\epsilon_{0}n_{2}}{2}FT\{\left| E_{L}\left(t,\xi \right) \right|^{2}E_{L}\left(t, \xi \right) \},
\end{split}
\label{eq2}
\end{equation}
where the subscripts $T$ and $L$ represent terahertz and laser, respectively. $E$ is the electric field. $\xi = z - v_{g}t$ is the coordinate moving at the group velocity $v_{g}$. $k\left(\omega\right) = n\left(\omega\right)\omega/c$ is the wave vector. For both $T$ and $L$ fields, absorption $\alpha$ and dispersion $D\left(\omega\right) = \omega\left[n\left(\omega\right)-n_{g}\left(\omega_{0}\right)\right]/c$ are considered. Terahertz free carrier absorption via three laser photon absorption is not taken into account because its expected coefficient, 13 cm$^{-1}$ \cite{zhong2015optimization}, is much smaller than terahertz single-photon absorption ($\sim$0.14 $\mu$m$^{-1}$ at 14.6 THz) at our laser intensity of 2.5 TW/cm$^{2}$. The last term in Eq. (\ref{eq1}) corresponds to optical rectification, a source for terahertz radiation. The effective nonlinear coefficient $d_{\text{eff}}$ is obtained from Ref. \cite{sussman1970tunable}. Also, cascading and self-phase modulation (SPM) effects are included in the third and fourth terms in Eq. (\ref{eq2}), respectively. The cascading term represents a terahertz-induced modulation on the laser pulse. In the SPM term, the nonlinear refractive index of $n_{2} = 10^{-6}$ cm$^{2}$/GW \cite{ravi2014limitations} is used, and $FT$ denotes Fourier transform. The initial laser pulse condition is assigned to be identical to the experimental one from Fig. \ref{fig1}(b) in all calculations. 

\begin{figure}[t!]
\centering
\includegraphics[width=10cm]{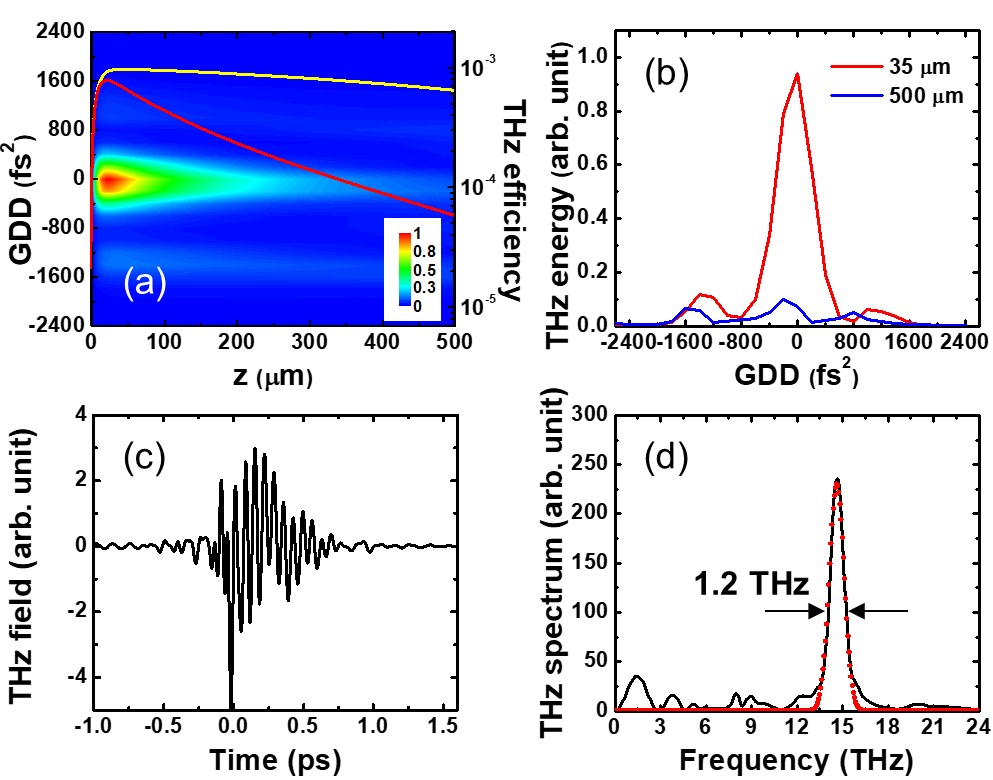}
\caption{(a) Simulated terahertz output energy (false color) as functions of crystal thickness and input laser GDD. Co-plotted is the terahertz conversion efficiency obtained with dispersion effect alone (yellow line) or with additional cascading and SPM effects (red line). (b) Simulated terahertz energy as a function of laser GDD at $z = $ 35 $\mu$m (red line) and $z = $ 500 $\mu$m (blue line). (c) Simulated terahertz waveform and (d) corresponding spectrum (black line) at $z = $ 23 $\mu$m with GDD $=$ 0 fs$^{2}$}. 
%\textcolor{red}{}}
\label{fig4}
\end{figure}

Figure \ref{fig4}(a) shows a simulation result illustrating terahertz energy observed at a distance $z$ from the crystal front surface ($z = 0$) for a range of initial laser GDD values. Here the input laser energy is 2 J with fluence of 0.1 J/cm$^{2}$. The terahertz energy is obtained by integrating the frequency components at 10--20 THz. The red solid line represents laser-to-terahertz conversion efficiency obtained  for GDD $=$ 0 fs$^{2}$. It shows an effective interaction length of $z_{\text{eff}} = $ 23 $\mu$m, providing a maximum conversion efficiency of $\sim$0.08\%. Here the laser field is assumed to be polarized along the extraordinary axis of cLN. Interestingly, with the cascading and SPM effects in Eq. (\ref{eq2}) excluded, the effective interaction length increases to $z_{\text{eff}} = $ 45 $\mu$m (yellow line) with the maximal conversion efficiency reaching 0.1\%. Here the cascading effect plays a negative role in terahertz generation. This is because it slowly modulates the laser intensity envelope, broadening the laser spectrum. Combined with dispersion, this stretches the laser pulse duration, consequently weakening the optical rectification process. This cascading effect also explains the decelerating conversion efficiency observed in Fig. \ref{fig3}(f) for laser energy of >1.8 J \cite{zhong2015optimization}. By contrast, the SPM effect is negligible in our condition because the expected nonlinear refractive index transient is as small as $\Delta n = $ 0.0025 at laser intensity of  2.5 TW/cm$^{2}$.

Figure \ref{fig4}(b) shows simulated terahertz output energy as a function of laser GDD obtained at $z = $ 35 $\mu$m (red solid line) and 500 $\mu$m (blue solid line). With the measured spectral intensity and phase taken into account, the GDD dependence is well reproduced, consistent with our observation shown in Fig. \ref{fig3}(a) and Fig. \ref{fig3}(c). Figure \ref{fig4}(c) shows a simulated terahertz waveform at $z = $ 23 $\mu$m. It clearly shows a multicycle terahertz pulse on top of slow terahertz oscillations (<3 THz), with the corresponding spectrum shown in Fig. \ref{fig4}(d). A Gaussian fit (red line) of the spectrum provides a narrow bandwidth of 1.2 THz in FWHM at a center frequency of 14.6 THz.

\section*{Conclusion}
In conclusion, we have demonstrated efficient multicycle terahertz pulse generation at 14.6 THz from bulk cLN crystals via phase-matched optical rectification of femtosecond Ti:Sapphire laser pulses with energy up to 2 J. Our experiments show that efficient terahertz generation can be achieved by using MgO-doped and/or thin cLNs. A thin (35-$\mu$m) cLN provides terahertz peak energy as high as 0.71 mJ even though its cut direction (y36-cut) is not optimized for terahertz generation. This suggests that much stronger terahertz pulses can be produced if one uses a LN crystal having a favorable cut (x or y), moderate MgO doping, and thickness close to the effective interaction length (23--150 $\mu$m). Also, a stoichiometric LN (sLN) crystal is another good choice because of its high nonlinearity and low terahertz absorption compared to cLNs \cite{palfalvi2005temperature, fujiwara1999comparison}. Those crystals will enable multi-mJ multicycle terahertz pulse generation via efficient (>0.1\%) optical rectification using $>$100 TW level lasers.

\section*{Funding}
Air Force Office of Scientific Research (FA9550- 25116-1-0163); Office of Naval Research (N00014-17-1-2705); GIST Research Institute (GRI 2020); Institute for Basic Science (IBS-R012-D1)

\end{document}